\documentclass[twocolumn]{article}
\usepackage{graphicx}
\usepackage{txfonts}
\begin{document}
\newcommand {\change} {\null}
\newcommand {\schange} {\null}
   \title{Collisional dissipation of Alfv\'{e}n waves in a
   partially ionised solar chromosphere}

   \author{J.E. Leake(1),          
          T.D. Arber(1), M.L. Khodachenko(2) \\
	 1. Space and astrophysics group,\\ 
	University of Warwick, \\
	Coventry, England CV4 7AL. \\ 
	J.E.Leake@warwick.ac.uk \\
         2. Space Research Institute, \\
              Austrian academy of sciences, \\
              A-8042 Graz,
              Austria. \\
              maxim.khodachenko@oeaw.ac.at
          }

  \date{Received by A\&A May 13, 2005; accepted July 27, 2005}
  \maketitle

  \abstract{Certain regions of the solar
     atmosphere are at sufficiently low temperatures to be 
     only partially ionised. The lower
     chromosphere contains neutral atoms, the existence of 
     which greatly increases the
     efficiency of the damping of waves due to collisional
     friction momentum transfer. More specifically
     the Cowling conductivity can be up to 12 orders of
     magnitude smaller than the Spitzer value, so that the
     main damping mechanism in this region is due to the collisions
     between neutrals and positive ions (\cite{khudachenko}).    
     Using values for the gas density and temperature as functions of
     height taken from the 
     VAL C model of the quiet Sun (\cite{vernazza}), an estimate is made
     for the dependance of the Cowling conductivity on height and
     strength of magnetic field. Using both analytic and numerical approaches the passage of
     Alfv\'{e}n waves over a wide spectrum through this partially
     ionised region is investigated. Estimates of the efficiency of
     this region in the damping of Alfv\'{e}n waves are made and
     compared for both approaches. We find that
     Alfv\'{e}n waves with
     frequencies above {\change 0.6 Hz} are completely damped and frequencies
     below 0.01 Hz unaffected. 
   }

   \maketitle
%%%%%%%%%%%%%%%%%%%%%%%%%%%%%%%%%%%%%%%%%%%%%%%%%%%%%%%%%%%%%%%%%%%%%%%%%%%
%%%%%%%%%%%%%%%%%%%%%%%%%%%%%%%%%%%%%%%%%%%%%%%%%%%%%%%%%%%%%%%%%%%%%%%%%%%
%Introduction________________________________________________________________

\section{Introduction}
MHD waves are vital phenomena in the study of the solar
atmosphere. They have been suggested as a source of heating of both
the chromosphere and the
corona (\cite{goodman1}, \cite{piddington}), based on the
damping of wave energy due to dissipation effects.

Three main energy dissipation mechanisms in the chromosphere 
have been studied in previous
work, namely collisional effects, viscous effects and thermal
conductivity effects.
Viscous damping of MHD waves is caused by the momentum transfer from
the thermal motion of particles. Collisional friction is caused by the
relative velocities of the species in the plasma. While thermal
conductivity is also related to kinetic effects and momentum transfer.
Khodachenko et al. (2004) and Khodachenko et al. (2005) showed that in the 
partially ionised plasma of the lower solar
chromosphere, collisional effects are far more important than the other
two. Also the presence of neutrals means that of all the collisions
to be considered it is the collisions between neutrals and ions that are
the most important component of this mechanism.

Alfv\'{e}n waves are a specific type of MHD wave.
Previous work on the generation and propagation of Alfv\'{e}n waves
has included the simulation of high frequency waves generated by
photospheric motions and the damping of these waves in the partially
ionised chromosphere above. This damping was suggested as 
an explanation of the formation of spicules {\change
(\cite{depontieu1}}, \cite{depontieu2}, \cite{james}).
Spicules are high, thin jet like structures consisting of
chromospheric plasma. They can reach heights of between 5000 and
9500 km whereupon they {\change either fall back or fade in the hot corona.} 
This work involved the numerical simulation of Alfv\'{e}n waves of
frequencies around 0.5 Hz damped by ion-neutral collisions 
in the WKB approximation. 
This approximation assumes that over a single wavelength there
is no spatial variation of variables. More specifically, the
wavelength itself does not change much on its own scale and the
damping scale is much larger than the wavelength.
De Pontieu et al. (2001) analytically calculated the damping 
time for Alfv\'{e}n waves of varying frequencies in 
model chromospheres of various solar structures, given estimates
for the plasma parameters in these regions. They estimated that
frequencies above about 0.1 Hz are unable to penetrate through to the
corona from the photosphere.

Other types of MHD waves in the solar atmosphere have also been
studied. Linear analysis of slow MHD waves in the frequency range
below $3.5 \times10^{-3}$ Hz 
has been used to show the damping as a possible heating mechanism for the
chromospheric network. 
The dissipation of the currents asscoiated with these waves, which is suggested to 
occur in thin
magnetic flux tubes with strengths of 700G to 1500G, heats the tubes and
hence the chromospheric network (\cite{goodman1}). 

The passage of waves through both the chromosphere and corona
has been studied using measurements of magnetic bright point (MBP) positions in the
photosphere. A frequency power spectrum for
horizontal motions (\cite{cranmer}) at the photospheric level 
were derived in the frequency range $10^{-5}$ to 0.1 Hz. 
Using these as a lower boundary condition,  a WKB approximation 
was applied to derive power spectra at different
heights in the atmosphere to show the effective damping of waves in
different regions. It was shown that
waves in this range ($10^{-5}$ to 0.1 Hz) may be evident up to a few solar
radii which suggests
that horizontal perturbations in this frequency range may be
unaffected by the partially ionised chromosphere.

Higher frequency waves have also been detected in the solar atmosphere.
Waves of frequencies up to 0.1Hz have been detected in the upper
chromosphere (\cite{deforest}) and transition region.
These waves were visible in the TRACE 1600 $\AA$ passband.
The exact source of these waves is uncertain and they are not
energetically significant to the chromosphere.
High frequency waves {\change (0.1 Hz)} in the lower chromosphere have
proved harder to find (\cite{fossum}). It has been suggested that high frequency 
Alfv\'{e}n waves in the
range 1 to 800 Hz can be created by small scale reconnection in the
chromospheric network, and their dissipation was considered as a possible heating 
mechanism for the
corona above (\cite{marsch}). 

Here we present both analytical estimates and full MHD simulations of the 
efficiency of damping of Alfv\'{e}n
waves in the partially ionised chromosphere. {\change We give
  estimates 
of the effective filter function for the damping of Alfv\'{e}n waves.}
This damping is assumed to be due to
the collisions of neutrals and positive ions in a model hydrogen
plasma. The Alfv\'{e}nic disturbances are assumed to be created at the
photospheric level by horizontal motions on the surface.

Given values for the plasma density and temperature in the lower
chromosphere taken from the
VAL C model of the quiet Sun (\cite{vernazza}), the Cowling conductivity is estimated.
Using the linear damping
approximation (\cite{braginskii}) the damping decrement for upwardly
travelling waves is calculated and the efficiency of this region in
damping the waves is estimated. 

Then full MHD numerical simulations of  
linear Alfv\'{e}n waves in a model
atmosphere which contains a partially ionised chromosphere are performed. 
We investigate the efficiency of collisional damping for a range of
frequency of wave. The
numerical results are compared with the theoretical estimates and
are shown to be in agreement with the linear damping approximation for
low {\change frequency} waves but differ for higher frequency waves.

%%%%%%%%%%%%%%%%%%%%%%%%%%%%%%%%%%%%%%%%%%%%%%%%%%%%%%%%%%%%%%%%%%%%%%
%%%%%%%%%%%%%%%%%%%%%%%%%%%%%%%%%%%%%%%%%%%%%%%%%%%%%%%%%%%%%%%%%%%%%%
%__________________________________________________________________
\section{Collisional friction damping in a partially ionised plasma}
To incorporate the damping of MHD waves due to ion-neutral collisions in a
partially ionised plasma, the generalized Ohm's law should be included
in the governing equations. (\cite{cowling}, \cite{braginskii}).
\begin{eqnarray}
\mathbf{E}+(\mathbf{v} \wedge \mathbf{B}) & = &
\frac{1}{\sigma} \frac{\nabla \wedge \mathbf{B}}{\mu_{0}} \\ 
& - & \frac{\xi^{2}}{\alpha_{n}}\left[\frac{\nabla \wedge \mathbf{B} \wedge
  \mathbf{B} \wedge \mathbf{B}}{\mu_{0}}\right]  \nonumber \\ 
& = &
 \frac{1}{\sigma}
 \mathbf{j}+\frac{\xi^{2}{B_{0}}^2}{\alpha_{n}}\mathbf{{j_{\bot}}}
\end{eqnarray}
where $\mathbf{v}$ is the fluid velocity, $\mathbf{B}$ is the magnetic
field,$\mathbf{E}$ is the electric
field, $\mathbf{j}=(\nabla \wedge \mathbf{B})/\mu_{0}$ is the current density and
$\mathbf{{j_{\bot}}}$ is the component perpendicular to the magnetic field.

The conductivity is defined by 
\begin{equation}
\sigma = \frac{n_{e} e^{2}}{m_{e}({\nu'}_{ei} + {\nu'}_{en})}
\end{equation}
and 
\begin{equation}
\alpha_{n} = m_{e}n_{e}{\nu'}_{en} + m_{i}n_{i}{\nu'}_{in}
\end{equation}
The number densities of species (ion, electron, neutral) are given
by $n_{i}, n_{e}, n_{n}$ respectively, and the masses by $m_{i},
m_{e}, m_{n}$.
\begin{equation}
\xi_{n}=\frac{m_{n}n_{n}}{m_{n}n_{n}+m_{i}n_{i}}
\end{equation}
is the relative
density of neutrals, and ${\nu'}_{ie}, {\nu'}_{ie}$ and
${\nu'}_{ie}$ are the effective collisional frequencies defined by
\begin{equation}
{\nu'}_{kl} = \frac{m_{l}}{m_{l}+m_{k}}\nu_{kl}
\end{equation}
where $k=e,i, l=i,n$.
Following the example of \cite{spitzer}, the collisional
frequencies of ions and electrons with neutrals are estimated by
\begin{eqnarray}
\nu_{in} & = & n_{n}\sqrt{\frac{8K_{B}T}{\pi m_{in}}}\Sigma_{in} \\
\nu_{en} & = & n_{n}\sqrt{\frac{8K_{B}T}{\pi m_{en}}}\Sigma_{en}.
\end{eqnarray}
where
\begin{equation}
m_{in} = \frac{m_{i}m_{n}}{m_{i}+m_{n}}.
\end{equation}
and $K_{B}$ is Boltzmann's constant.
The collisional frequency of electrons and ions is given by
\begin{eqnarray}
\nu_{ei} & = & 5.89.10^{-24}\frac{n_{i}\Lambda Z^{2}}{T^{3/2}}
\end{eqnarray}
(\cite{spitzer}).
$\Sigma_{in} = 5.10^{-19} \textrm{m}^{2}, \Sigma_{en} = 10^{-19}
\textrm{m}^{2}$ are the ion-neutral
and electron-neutral cross-sections respectively.
Here $\Lambda$ is the Coulomb logarithm.

{\change In the generalised Ohm's law (equations 1 and 2) we have
  neglected the pressure term, as the chromospheric plasma is
  relatively cold, and the Hall term. The Hall term can be dropped
  from the generalised Ohm's law if the plasma is magnetised, i.e. if
  the ions and electrons are tightly bound to the magnetic field. In
  terms of the ion-gyrofrequency and the collision time this condition
  can be written as 
\begin{equation}
\omega_{i} \tau \gg 1.
\end{equation}
Using the definition of the ion-neutral collision frequency this is
equivilant to
\begin{equation}
\frac{eB}{m_{i}}\sqrt{\frac{\pi
  m_{i}}{16k_{B}T}}\frac{1}{n_{n}\Sigma_{in}} \gg 1
\end{equation}
{\schange which depends on the strength of the magnetic field as well as the 
temperature and number density. Using values for temperature and number density taken 
from the 
VALC model of the Sun 
(\cite{vernazza}), and various models of magnetic field strengths in the chromosphere 
(see figure 1), 
this condition is  satisfied everywhere in the chromosphere.}
Khodachenko and Zaitsev (2001) showed that this condition does not
hold in the upper photosphere where the plasma is more weakly ionised.
In section 3 we will show that the damping mechanism we are
considering is not effective in the photosphere, and thus we only study
Alfv\'{e}n waves in the chromosphere for which the neglect of the Hall
term is valid.}

If the plasma is assumed to be entirely composed of hydrogen then the expressions
can be simplified by taking 
$m_{in}=m_{n}/2$, $\xi_{n}=\rho_{n}/\rho$ resulting in
\begin{equation}
\alpha_{n}=\frac{1}{2}\xi_{n}(1-\xi_{n})\frac{\rho^{2}}{m_{n}}\sqrt{\frac{16k_{B}T}{\pi 
m_{i}}}\Sigma_{in}.
\end{equation}

Using the definition for the Cowling conductivity as 
\begin{equation}
\sigma_{c}=\frac{\sigma}{1+\frac{{\xi_{n}}^2{B_{0}}^2\sigma}{\alpha_n}}
\end{equation}
and defining the Coulomb and Cowling resistivity as $\eta=1/\sigma$ and $\eta_{c}=1/\sigma_{c}$
respectively it is trivial to show that
\begin{equation}
\frac{{\xi_{n}}^2{B_{0}}^2}{\alpha_{n}} = \eta_{c} - \eta
\end{equation}
Thus equation 2 can be be written as
\begin{eqnarray}
\mathbf{E}+(\mathbf{v} \wedge \mathbf{B}) & = &
\eta \mathbf{j} + (\eta_{c}-\eta)\mathbf{j_{\bot}} \nonumber \\ 
& = & \eta \mathbf{{j_{\|}}}
+ \eta_{c} \mathbf{{j_{\bot}}}
\end{eqnarray}
and the frictional Joule heating term for this plasma
(\cite{braginskii}) is then given by
\begin{eqnarray}
Q & = & (\mathbf{E}+(\mathbf{v} \wedge \mathbf{B})).\mathbf{j}
\nonumber \\
  & = & \eta{{j_{\|}}^2} +\eta_{c}{{j_{\bot}}^2} 
\end{eqnarray}
with $\mathbf{j_{\bot}}$ and $\mathbf{j_{\|}}$ being the components
of $\mathbf{j}$ perpendicular and parallel to $\mathbf{B}$.
Thus in a partially ionised plasma the Coulomb resistivity acts to diffuse
currents parallel to the magnetic field and the Cowling resistivity
acts to diffuse currents perpendicular to the magnetic field.
The damping decrement ($\delta$) for this effect was derived by
Braginskii (1965)
for Alfv\'{e}n waves propagating with angular frequency $\omega$ and
wavenumber $\mathbf{k}$ and is given by  
\begin{equation}
\frac{1}{\tau} = 2 \omega \delta_{Joule} =  \frac{1}{\mu_{0}}\eta {k_{\bot}}^2 +
\frac{1}{\mu_{0}}\eta_{c}{k_{\|}}^2  
\end{equation}
Khodachenko et al. (2004, 2005) showed that for typical magnetic field
strengths and using the VAL C model for the plasma parameters,
the ratio of Coulomb to Cowling resistivity can vary from 
1 at the solar surface (z=0) down to $10^{-12}$ at heights of 2000 km. For the
partially ionised chromosphere
\begin{equation}
\frac{\eta}{\eta_{c}} = \frac{1}{1+{\xi_{n}}^2 \frac{{B_{0}}^2}{\eta
    \alpha_{n}}} << 1.
\end{equation}
Hence when considering frictional damping mechanisms, the ion-neutral
collisions are by far the most important.

%%%%%%%%%%%%%%%%%%%%%%%%%%%%%%%%%%%%%%%%%%%%%%%%%%%%%%%%%%%%%%%%%%%%%%%%
%%%%%%%%%%%%%%%%%%%%%%%%%%%%%%%%%%%%%%%%%%%%%%%%%%%%%%%%%%%%%%%%%%%%%%%%
\section{Analytic approach}

In order to evaluate the expression for the COwling conductivity $\eta_{c}$
an estimate for the neutral fraction $\xi_{n}$ is required (see equation 15). Although the
VAL C model gives us the ionisation degree for the Sun our model is a
hydrogen plasma. Thus $\xi_{n}$ is calculated from the density and
temperature values only of the VAL C model. 
Here we follow the method of De Pontieu (1999), assuming a hydrogen
plasma where the number of electrons is equal to the number of protons
($n_{i}=n_{e})$.

The solar chromosphere, and spicules especially are not in
LTE. Hence a simple one-level model for the hydrogen atom is inadequate
for these conditions (\cite{pottasch}).
A two-level model is used instead for the hydrogen atom, so that the
ionisation equation (\cite{brown}) is
\begin{equation}
n\frac{\partial x}{\partial t} =
\sum(C_{j}+P_{j}+{C_{j}}^{*}+{P_{j}}^{*})
\end{equation}
where \emph{n} is local number density, and $x$ is the {\change ionisation degree.}
 $C_{j}$ and $P_{j}$ are the ionisation rates from level \emph{j} due to
thermal collisions and local radiation field respectively.
${C_{j}}^{*}$ and ${P_{j}}^{*}$ are the ionisation rates by non-thermal
particles and external radiation field.
In the first approximation photoionisation from level 2 to level 1 is
provided by the external field alone, so that
$P_{1}={P_{1}}^{*}=0$. This external field is simply the photospheric
radiation field at temperature $T_{R}$, suitably diluted by factor
$w$.
The ionisation equation can be further simplified by noting that thermal
collision ionisation is unimportant when compared to
photoionisation (\cite{ambartsumyan}). Thus the ionisation equation is
balanced by photo-ionisation from level 2 to level 1 and spontaneous
recombination for the return route.
\begin{equation}
n\frac{\partial x}{\partial t} = {P_{2_{+}}}^{*} - P_{2-}
\end{equation}
The steady state solution to this equation is given by (\cite{thomas})
\begin{eqnarray}
\frac{{n_{i}}^2}{n_{n}} & = & \frac{f(T)}{b(T)} \\
f(T) & = & \frac{(2\pi
  m_{e}K_{B}T)^{\frac{3}{2}}}{h^{3}}\exp({-\frac{X_{i}}{K_BT}}) \\
b(T) & = & \frac{T}{wT_{R}}\exp \left[\frac{X_{i}}{4K_{B}T}(\frac{T}{T_{R}}-1
) \right]
\end{eqnarray}
where $T_{R}$ is the temperature of the photospheric radiation field 
and $w=0.5$ is its dilution factor.

Using the solar atmospheric plasma variables given by the VAL C model
the neutral fraction $\xi_{n}$ and the 
profile of $\eta_{c}$ with height can be calculated for a given 
magnetic field profile.

As we intend to simulate the motion of waves in the chromospheric
region of the solar atmosphere, we choose as our magnetic field model
a spreading vertical flux tube representing an open magnetic structure 
with its {\change footpoint} in the photosphere. As Alfv\'{e}n waves propagate
along a fieldline we can restrict ourselves to one spatial dimension
and need only define the variation of the magnetic field 
 as a function of distance along the fieldline. {\change Thus the
   horizontal expansion of the tube does not enter explicitly into our
   calculations. We choose a magnetic field profile to reflect the
   conservation of flux as the tube expands.} A power law dependance
 on density is chosen 
\begin{equation}
B = B_{0} \left({\frac{\rho}{\rho_{ph}} } \right)^{\alpha}
\end{equation}
where $\rho_{ph}$ is the photospheric density $2.7 \times 10^{-4}
\textrm{kg} / \textrm{m}^{3}$ and $B_{0}$ is 1200G.
This profile is chosen so that at the base of the domain the magnetic
 field is approximately 1000G and falls to a value of about 10G at a height
 of 3000km. This is a simplistic model for the magnetic field but
 captures the decrease with height and enables comparison of analytic
 and numerical results. {\change Martinez-Pillet et al (1997) showed
   that the field strength in flux tubes can drop to as little as 300G
   at heights of only 300km so we use this as the minimum value of
   magnetic field at this height for our models.}
The magnetic field profile for four different
 power law profiles are shown in figure
 1. A typical calulated value of $\eta_{c}$ as a function of height
 for the case when $\alpha=0.3$ is
 shown in figure 2. Khodachenko et al. (2005) calculated
 $\frac{\eta}{\eta_{c}}$ for the solar atmosphere given constant
 magnetic fields. The ratio at a height of 1500 km can be as low as
 $10^{-6}$ giving a value of the Coulomb resistivity of $8 \times
 10^{-5} \textrm{m}^{2}/ \textrm{s}$. 

%_____________________________________________________________
%                 A figure as large as the width of the column
%-------------------------------------------------------------
   \begin{figure}
   \begin{center}
   \includegraphics[width=7cm]{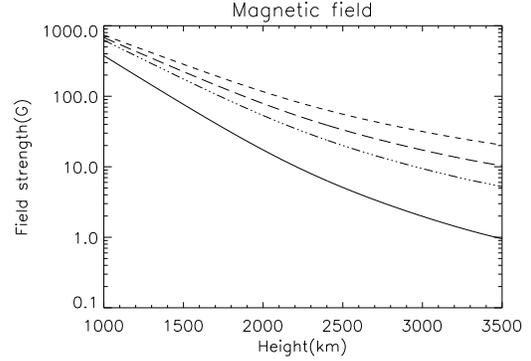}
      \caption{The variation of the model background magnetic field 
        with height. The four curves relate to the value of the
        exponent $\alpha$ = 0.3 (small dash line), 0.35 (big dash line),
        0.4 (dot-dash line) and 0.6 (solid line). }
             
   \end{center}
   \end{figure}
%_____________________________________________________________
%                 A figure as large as the width of the column
%-------------------------------------------------------------
   \begin{figure}
   \begin{center}
   \includegraphics[width=7cm]{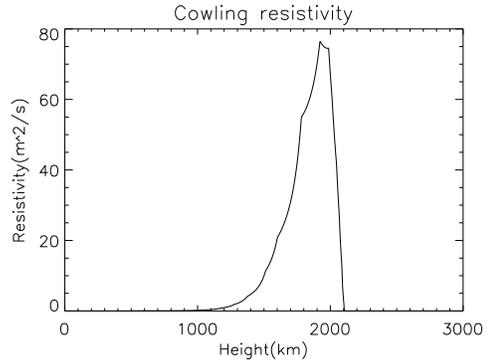}
      \caption{Cowling resistivity as a function of height calculated
        from VAL C model of the quiet Sun and varying $B_{0}$ with
        height according to equation 23 with $\alpha = 0.3$.}
             
   \end{center}
   \end{figure}

Equation 18 gives the inverse damping timescale of Alfv\'{e}n waves with angular
frequency $\omega$ and wavenumber $k_{||}$ along the magnetic field, 
propagating through a region of constant $\eta_{c}$.
\begin{equation}
\frac{1}{\tau} = 2 \omega \delta = \frac{\eta_{c}}{\mu_{0}}{k_{||}}^{2}
\end{equation}

The linear MHD analysis used to obtain damping
decrements (\cite{braginskii}) assumes that the
damping decrement $\delta$ is much less than 1. This implies that
\begin{equation}
\delta = \frac{\eta_{c}{k_{\|}}^2}{2\mu_{0}\omega} \ll 1
\end{equation}
or that
\begin{equation}
\omega \ll \omega_{crit} = \frac{2 \mu_{0} {c_{A}}^2}{\eta_{c}}.
\end{equation}

For the model used here both $\eta_{c}$ and the Alfv\'{e}n
speed
\begin{equation}
c_{A} = \frac{B}{ ({ \mu_{0} \rho})^{\frac{1}{2}} }
\end{equation}
are functions of height so the damping
decrement also varies with height.
Over a small timestep \emph{dt} the amplitude of the propagating wave changes by
\begin{eqnarray}
A & = & A_{0}\exp\left(-\frac{dt}{\tau} \right) \\
  & = & A_{0}\exp\left(-dt\frac{\eta_{c}(z)}{\mu_{0}}{k_{||}(z)}^2 \right)
  \nonumber \\
  & = &
  A_{0}\exp\left(-dz\frac{\eta_{c}(z)}{\mu_{0}{c_{A}(z)}^3}{\omega}^2
  \right) \nonumber 
\end{eqnarray}
By integrating over the partially ionised region where $\eta_{c}$ is
non-zero (figure 2) we get an estimate for the total change in amplitude due to
the damping mechanism
\begin{equation}
E = \frac{A_{0}-A}{A_{0}} = 1
-exp\left[- \omega^2 \int{\frac{\eta_{c}(z)}{\mu_{0}{c_{A}(z)}^3}dz} \right]
\end{equation}
where the integral can be performed numerically. The results from this 
approximation are shown in figure 3 for the four different magnetic
field models (different $\alpha$).
From these estimates it appears that in the absence of any other
damping mechanisms and stratification effects, high frequency
Alfv\'{e}n waves are unable to pass through this partially 
ionised region without being completely damped. For the magnetic
field model with $\alpha$ = 0.3 this can be as low as 0.1 Hz. 

%_____________________________________________________________
%                 A figure as large as the width of the column
%-------------------------------------------------------------
   \begin{figure}
   \begin{center}
   \includegraphics[width=7cm]{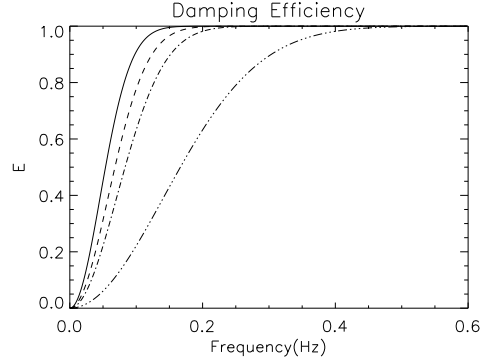}
      \caption{Analytical estimates for the damping efficiency of the
        partially ionised region of the solar chromosphere
        as a function of frequency. The four lines represent the
        four magnetic field profile given by $\alpha = 0.3$
        (solid line). 0.35 (dash line),  0.4 (dot-dash line) and
      0.6 (triple dot-dash line).}
             
   \end{center}
   \end{figure}

%%%%%%%%%%%%%%%%%%%%%%%%%%%%%%%%%%%%%%%%%%%%%%%%%%%%%%%%%%%%%%%%%%%%%%%%
%%%%%%%%%%%%%%%%%%%%%%%%%%%%%%%%%%%%%%%%%%%%%%%%%%%%%%%%%%%%%%%%%%%%%%%%
\section{Numerical method}
The propagation of Alfv\'{e}n waves in a partially ionised plasma is
modelled in 1D by numerically solving the non-ideal MHD equations, given here in
dimensionless form. 
\begin{eqnarray}
\frac{\partial \rho}{\partial t} & = & -\nabla .(\rho
\mathbf{v}) \\
\frac{\partial}{\partial t}(\rho \mathbf{v}) & = & -\nabla
. (\rho \mathbf{vv}) + (\mathbf{j} \wedge \mathbf{B}) - \nabla P +
\rho \mathbf{g} \\
\frac{\partial \mathbf{B}}{\partial t} & = & \nabla \wedge
(\mathbf{v} \wedge \mathbf{B}) - \nabla \wedge (\eta \mathbf{j_{\|}})
\nonumber \\
& - & \nabla \wedge (\eta_{c} \mathbf{j_{\bot}}) \\ 
\frac{\partial}{\partial t}(\rho \epsilon) & = & -\nabla .(\rho
\epsilon \mathbf{v}) - P\nabla .\mathbf{v} \nonumber \\ 
& + & \eta {j_{par}}^{2} + \eta_{c}{j_{perp}}^{2} 
\end{eqnarray} 
which also include the Coulomb and Cowling resistivities
$ \eta = 1/\sigma $ and $\eta_{c} = 1/\sigma_{c} $.
$\mathbf{B}$ is the magnetic field, $ \mathbf{j}$ is the current
density,  $\mathbf{v}$ is the
velocity, P is the thermal pressure, and $\epsilon = 
P/(\gamma -1)\rho $ is the specific energy density.
The constants $\mu_{0}$ and $\gamma$ have the standard meanings of magnetic permittivity of
free space and the ratio of specific heats respectively. Although for
a partially ionised plasma $\gamma < 5/3$ we found the value had
little effect on the results so $5/3$ was used in all simulations.

It is worth noting that no viscous effects are included in these
model equations as only collisional damping is being investigated in
this work.

The simulations are carried out using a 2D MHD shock capturing code 
(Lare2d), applied in 1D only. It uses a staggered grid with density, 
pressure and specific energy density defined in the cell centres, magnetic field at
the cell faces and velocities at the vertices. For further details see
Arber et al. (2001).
The code is adapted so that the resistive term in the induction
equation updates the perpendicular and parallel components of current
density separately, according to equation 16.
Although the code is 1D all vector variables have 3 components.

The simulation domain extends 3000 km vertically, while the computational grid
consists of 2000 cells. The
number of cells vertically is restricted by the need to resolve the
smallest wavelengths in the parametric study. The vertical boundary
conditions are line-tied which are perfectly reflecting. Thus the 
simulation ends when the Alfv\'{e}n waves have reached the upper boundary.

Using the VAL C model we construct the temperature profile, and the
density is determined by solving the hydrostatic equilibrium equation.
\begin{equation}
\frac{\partial P}{\partial z} = -\rho \mathbf{g}
\end{equation}
The background equilibrium is shown in figure 4.
%_____________________________________________________________
%                 A figure as large as the width of the column
%-------------------------------------------------------------
   \begin{figure}
   \begin{center}
   \includegraphics[width=7cm]{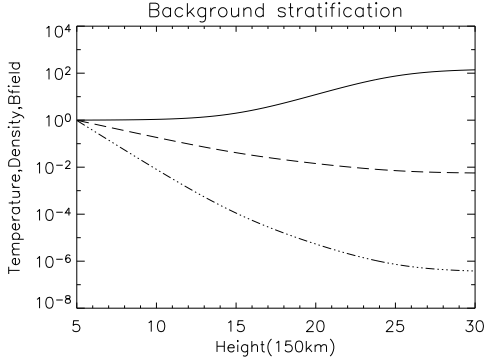}
      \caption{The background plasma temperature (solid line),
        density(dot-dash) and magnetic field (dash), where
        $\alpha=0.3$. The values are all
        given in dimensionless units.}             
  
   \end{center}
   \end{figure}

All quanities were non-dimensionalised by using values defined at z=0 
from the VAL C model, thus 
\begin{eqnarray}
r^{*} & = & 150 km  \\
\rho^{*} & = & 2.7 \times 10^{-4} kg/m^{3} \nonumber\\ 
v^{*} & = & 6515 m/s \nonumber \\ 
t^{*} & = & 23 s \nonumber \\ 
T^{*} & = & 6420 K \nonumber \\ 
P^{*} & = & 1.17 \times 10^{4} Pa \nonumber \\
B^{*} & = & 1000 G \nonumber 
\end{eqnarray}
Conversion to variables with dimensions merely requires the
multiplication of internal variables by these values.

We model linear propagating Alfv\'{e}n waves by driving the
horizontal velocity $v_{x}$ continuously at the bottom of the domain 
with a sinusoidal driving function, the amplitude of which 
did not affect the results. {\change Typical driving velocities were 600 $m/s$
and a range of 80 to 800 $m/s$ was tested.}

As already mentioned in section 2 the Cowling resistivity 
$\mathbf{\eta_{c}}$ is much
larger than the Coulomb resistivity $\mathbf{\eta}$ at chromospheric 
heights. In fact
over the domain being simulated, the calculated value of $\eta$ using
VALC values for the plasma never exceeds the
value of numerical roundoff in the code, and hence all simulations are
run with $\eta=0$.

The dissipation of these waves can then be investigated by observing the
change in amplitudes of the perturbations. 
Figure 5 shows a typical profile of the perpendicular velocity and
magnetic field
perturbations with height for a driving frequency of
0.07 Hz, for the case when no damping mechanism is applied and when
collisonal friction damping is applied with the given profile of
$\eta_{c}$ (figure 2). As in figure 4 the value of $\alpha$ is 0.3.

%_____________________________________________________________
%                 A figure as large as the width of the column
%-------------------------------------------------------------
   \begin{figure}
   \begin{center}
   \includegraphics[width=3cm]{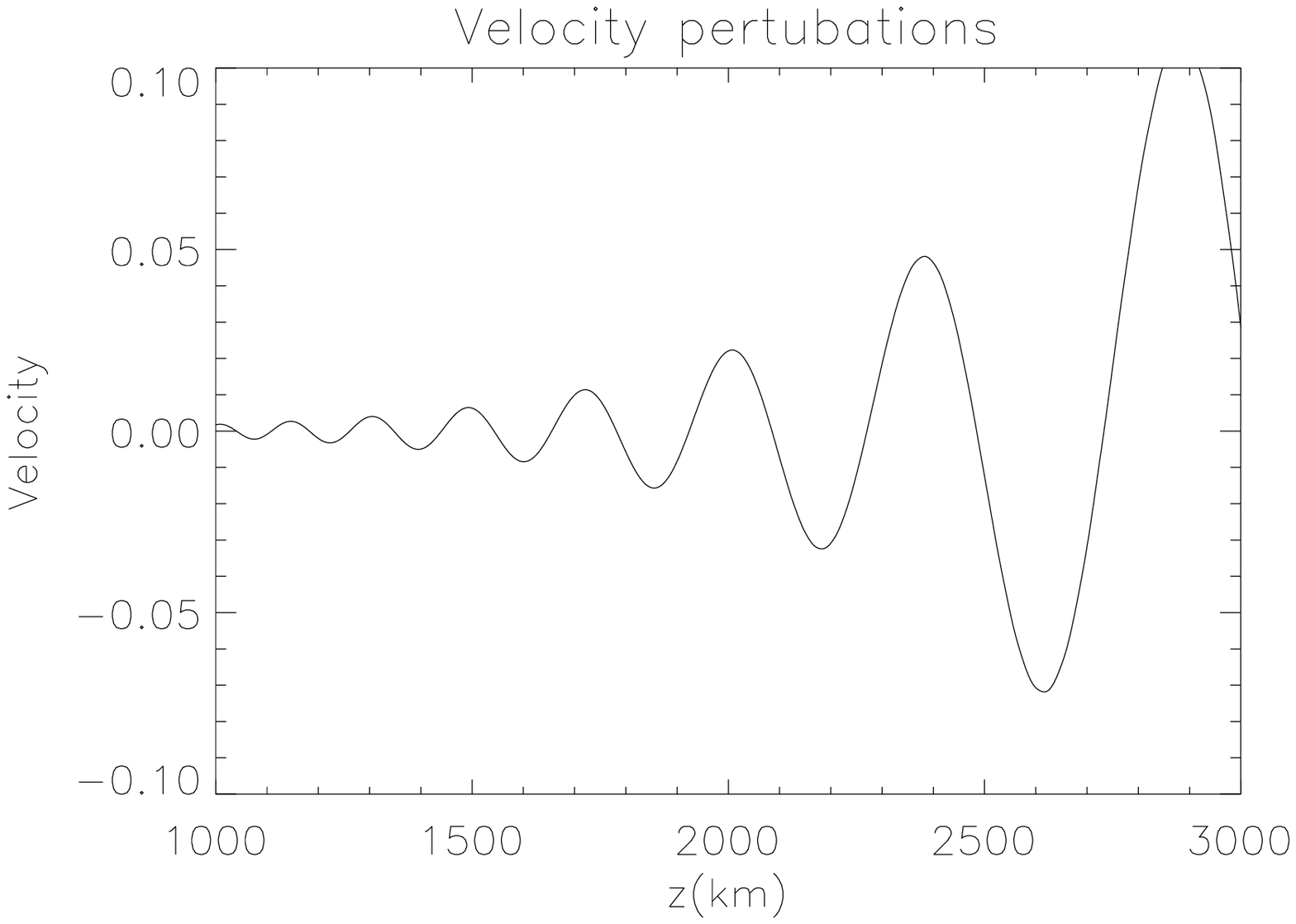}
   \includegraphics[width=3cm]{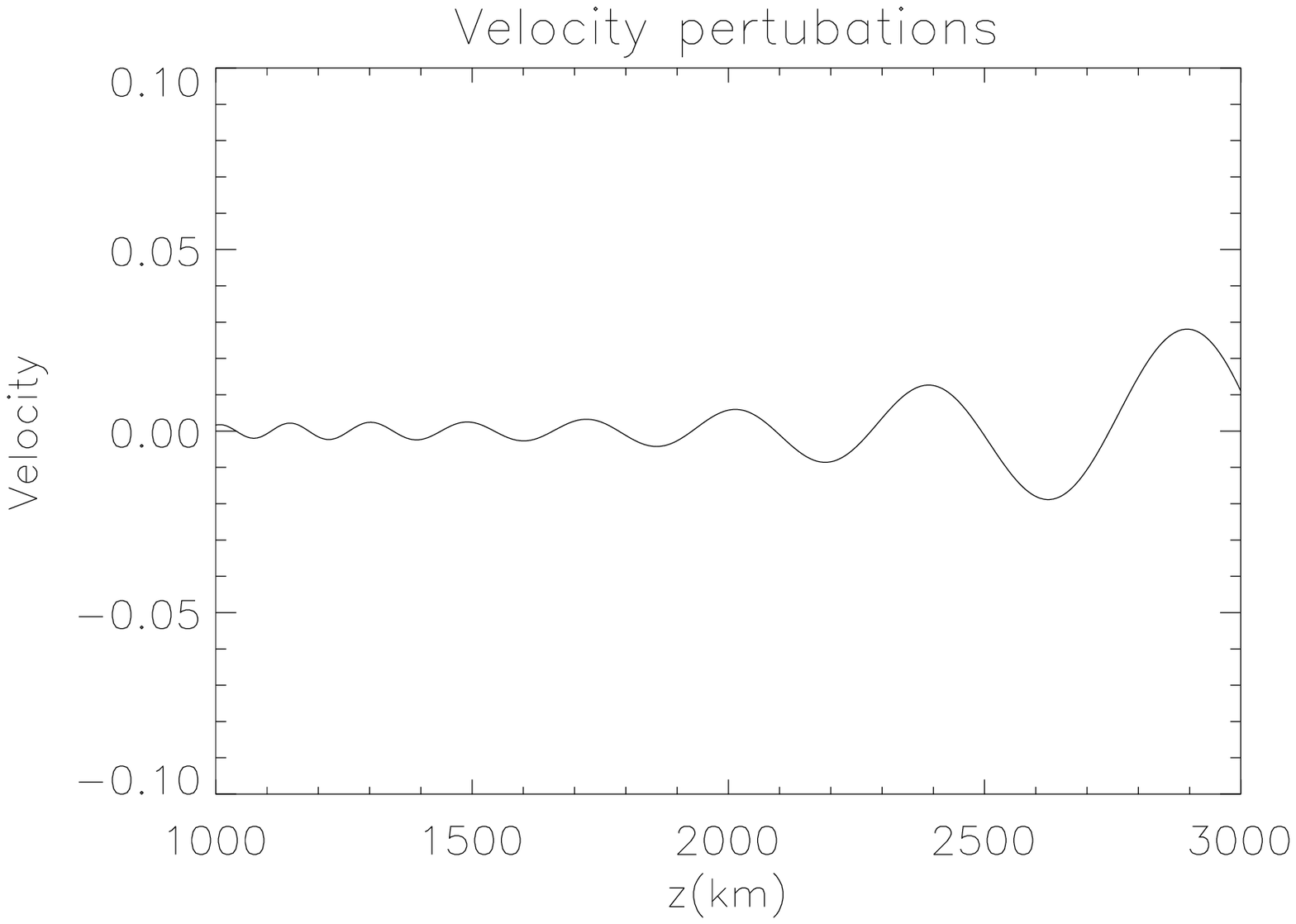}
   \includegraphics[width=3cm]{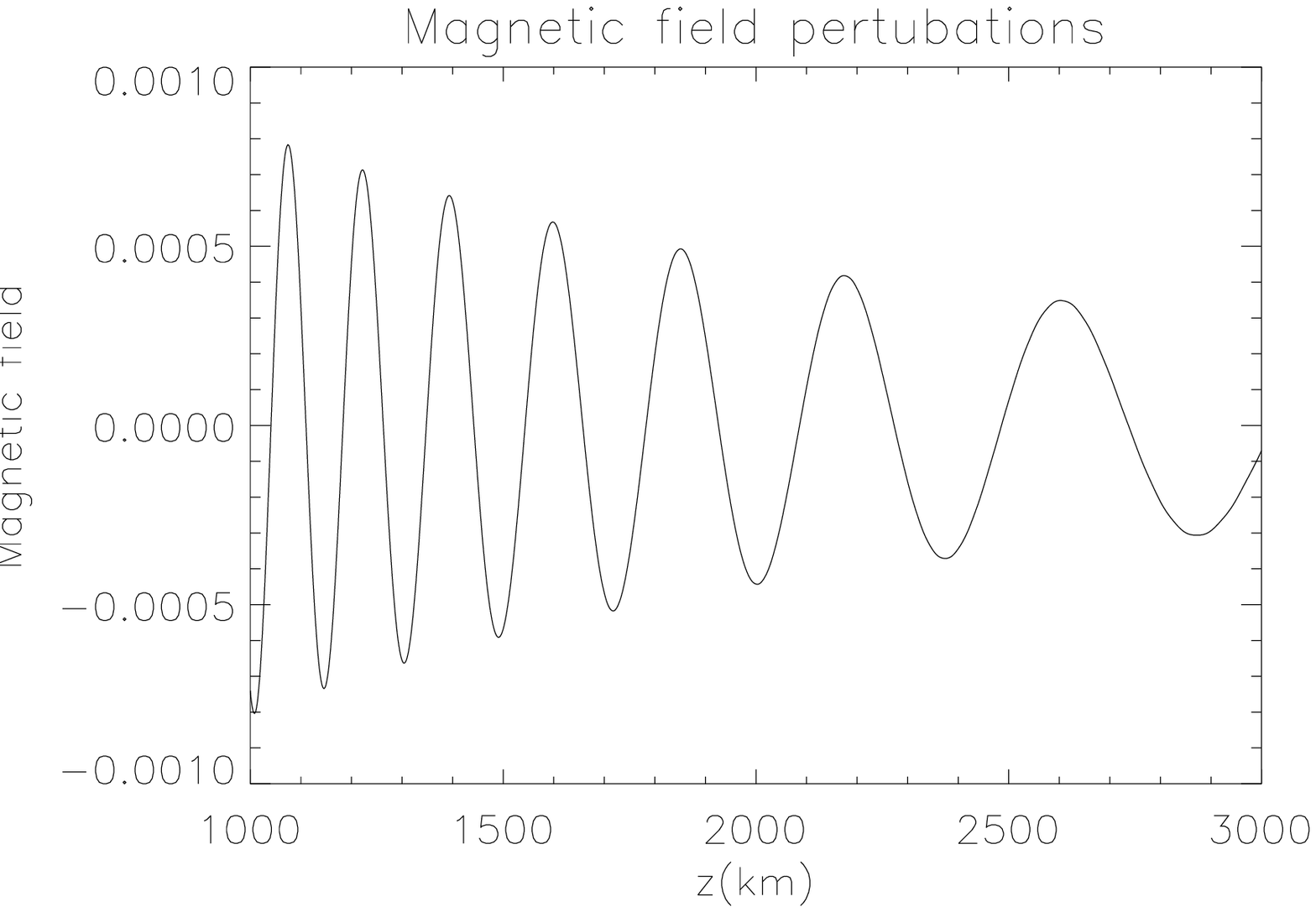}
   \includegraphics[width=3cm]{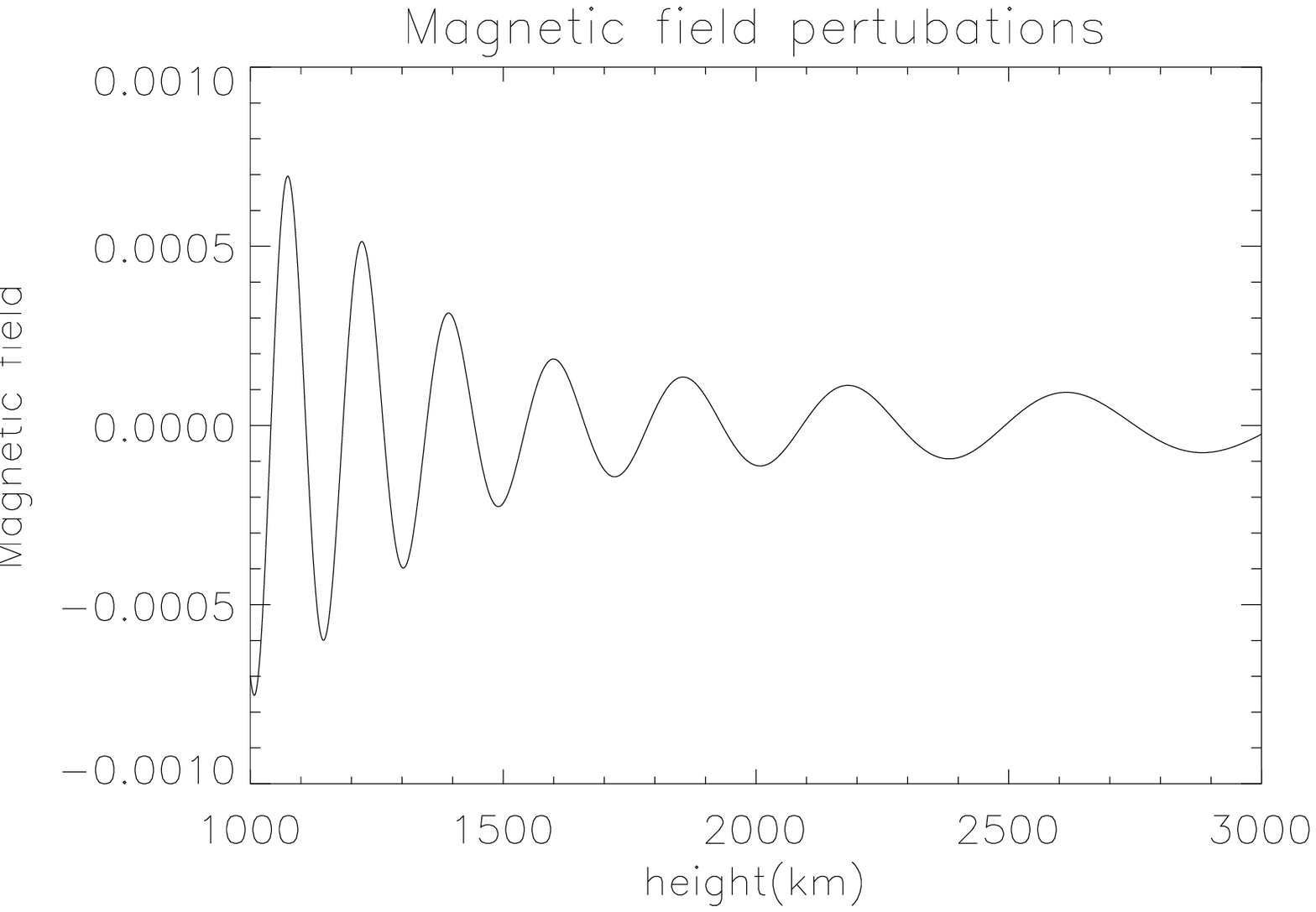}
   \caption{Velocity (top slides) and magnetic field (bottom slides) 
     perturbations against height
   for a typical run with $f=0.07Hz$, The left hand slides are for the
   case of $\eta_{c}=0$ and the right hand slides are for the 
   $\eta_{c}$ profile given by figure 2. The velocity and magnetic
   field are given in dimensionless units}
       
   \end{center}
   \end{figure}

By comparing the propagation of Alfv\'{e}n waves in the presence of a
partially ionised region and without, we can study the efficiency of
damping due to collsional friction of ions and neutrals in the
partially ionised chromosphere.

As can be seen from figure 5, the gravitational stratification causes an
increase in amplitude of the velocity pertubation and a
decay in the corresponding magnetic field pertubations. This is
a consequence of energy conservation in the travelling wave. For a
rigorous mathematical treatment of this see Moortel and Hood (2004).
This makes a direct measuremnt of the efficiency of the collisional
frictional damping non-trivial.

{\change In order to obtain an estimate for the efficiency of collisional
friction damping we compare the Poynting flux carried by the waves in
the case when no damping mechanism is present and when there is
damping due to ion-neutral collisions. The time-averaged Poynting flux for the
Alfv\'{e}n waves is given by
\begin{equation}
<S> = \frac{B_{0}<b_{x}v_{x}>}{\mu_{0}}
\end{equation}
where $B_{0}$ is the vertical background magnetic field and $b_{x}$, $v_{x}$
are the perturbations in magnetic field and velocity.
By calculating the ratio of Poynting flux at a height above the
partially ionised region of the model chromosphere we can estimate the
efficiency of the damping mechanism. The ratio of damped to undamped
Poynting flux at a height of 2500km above the surface (i.e. above the
region of high Cowling conductivity) is given by
\begin{equation}
\gamma = \frac{<S>_{damped}}{<S>_{undamped}}
\end{equation}
and for direct comparison with analyic results the efficiency of
damping is estimated simply by
\begin{equation}
E_{num} = 1-\gamma.
\end{equation}

Figures 6, 7, 8 and 9 show this estimate as a function of frequency, along with 
the estimate from the analytic
approach (solid lines). The four different plots are for the four
different magnetic field profiles, with $\alpha$ being 0.3, 0.35, 0.4
and 0.6 respectively.
As the power law for the magnetic field changes so does our estimate
for the damping efficiency, as the value of $\eta_{c}$ is dependant on
$|\mathbf{B}|$.

%_____________________________________________________________
%                 A figure as large as the width of the column
%-------------------------------------------------------------
   \begin{figure}
   \begin{center}
   \includegraphics[width=7cm]{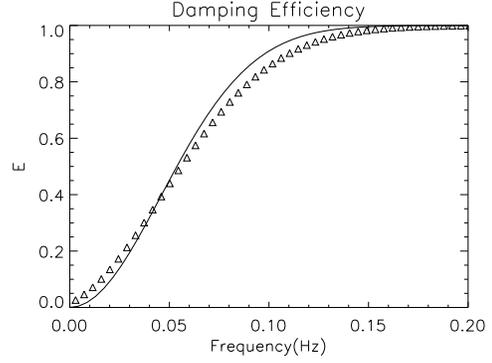}
      \caption{Estimates of the efficiency of damping
      due to ion-neutral collisions for the magnetic field profile
      with power law $\alpha =0.3$. The solid line is the estimate
      due to analytic approches
       and the triangles represent the estimates obtained from
      numerical data.}             
         
   \end{center}
   \end{figure}

%_____________________________________________________________
%                 A figure as large as the width of the column
%-------------------------------------------------------------
   \begin{figure}
   \begin{center}
   \includegraphics[width=7cm]{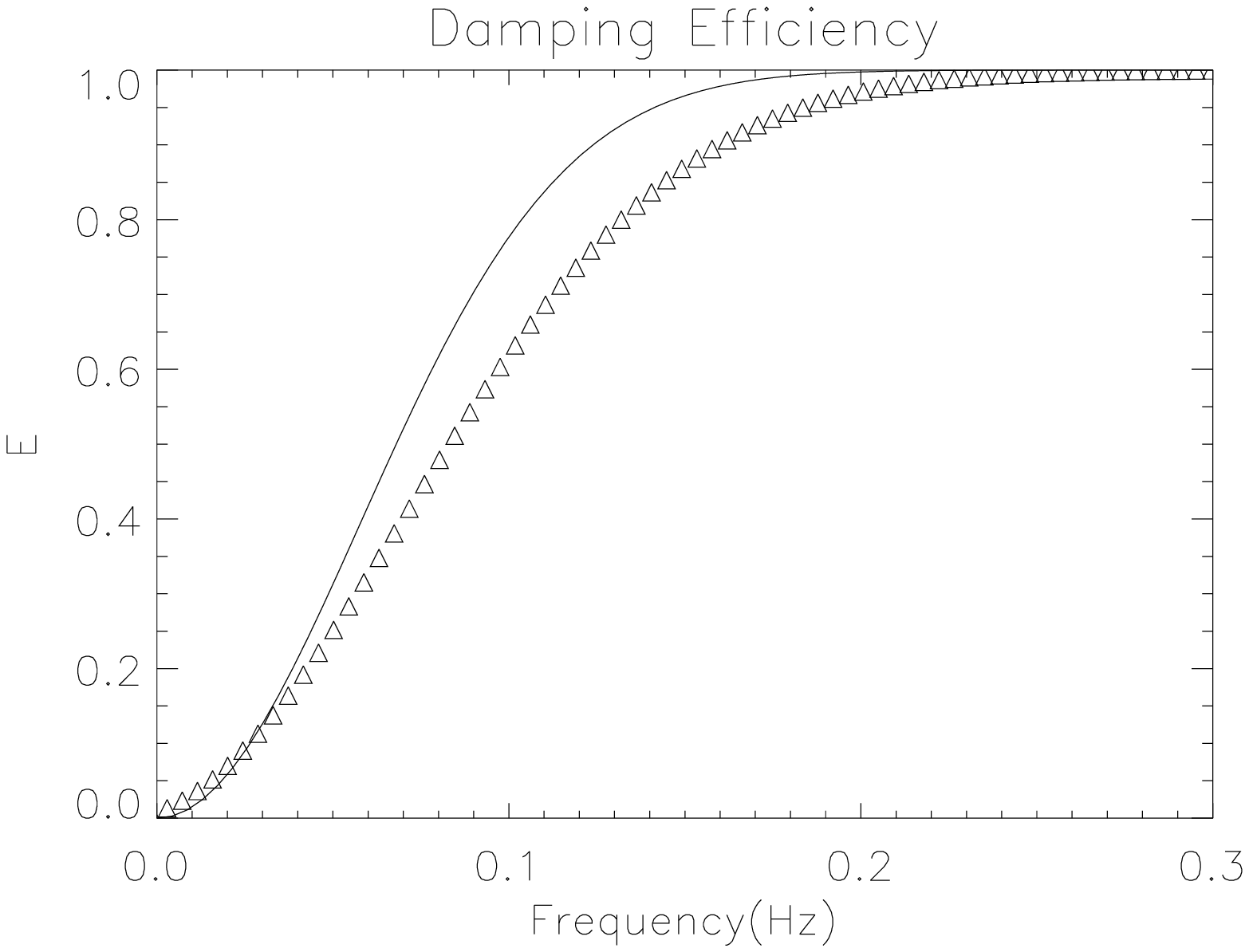}
      \caption{Estimates of the efficiency of damping
      due to ion-neutral collisions for the magnetic field profile
      with power law $\alpha = 0.35$. The solid line is the estimate
      due to analytic approches
       and the triangles represent the estimates obtained from
      numerical data.}             
         \label{amp2}
   \end{center}
   \end{figure}

%_____________________________________________________________
%                 A figure as large as the width of the column
%-------------------------------------------------------------
   \begin{figure}
   \begin{center}
   \includegraphics[width=7cm]{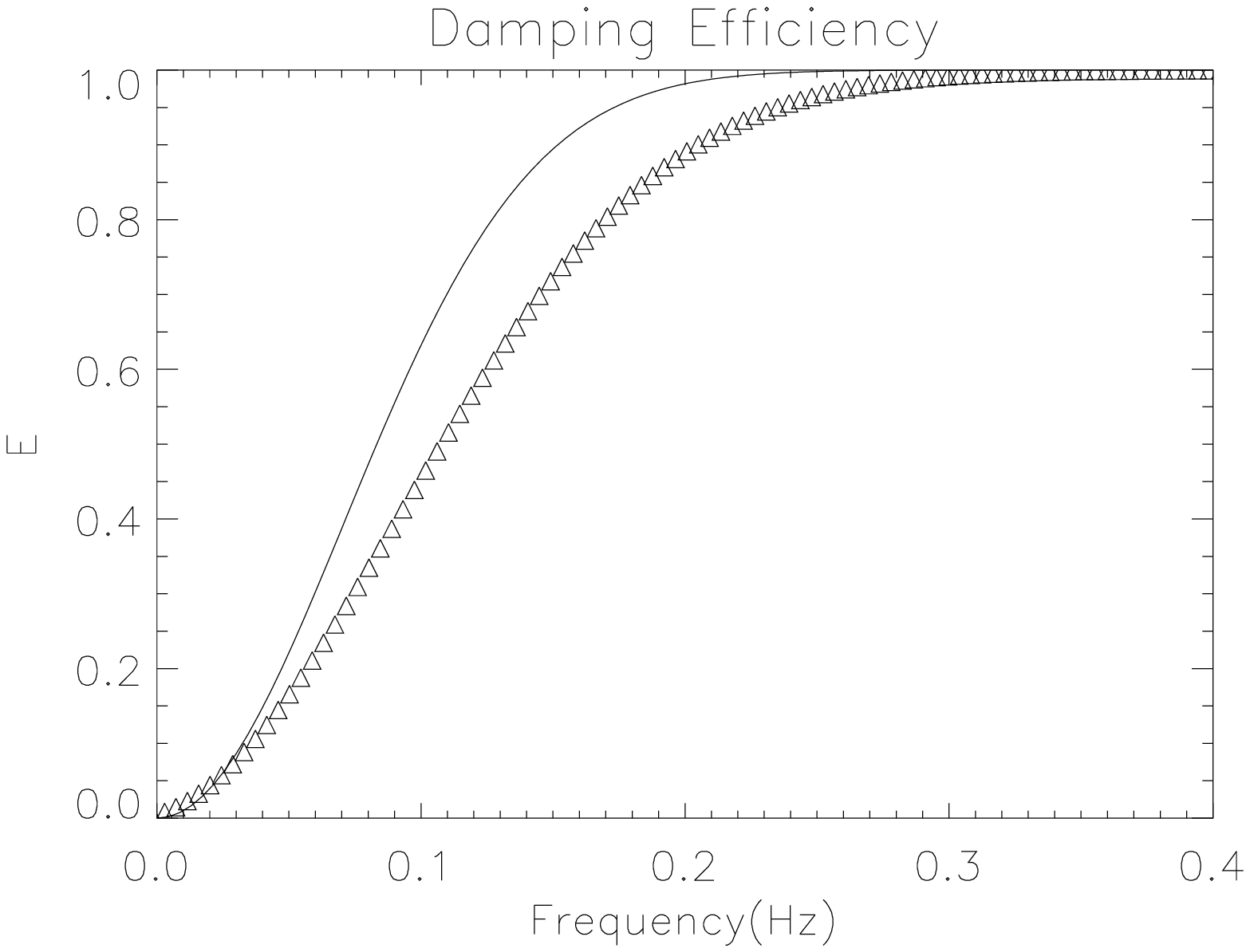}
      \caption{Estimates of the efficiency of damping
      due to ion-neutral collisions for the magnetic field profile
      with power law $\alpha = 0.4$. The solid line is the estimate
      due to analytic approches
       and the triangles represent the estimates obtained from
      numerical data.}             
         
   \end{center}
   \end{figure}

%_____________________________________________________________
%                 A figure as large as the width of the column
%-------------------------------------------------------------
   \begin{figure}
   \begin{center}
   \includegraphics[width=7cm]{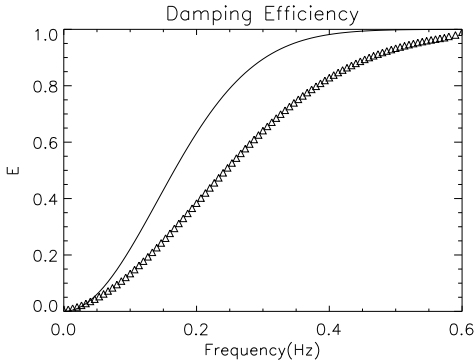}
      \caption{Estimates of the efficiency of damping
      due to ion-neutral collisions for the magnetic field profile
      with power law $\alpha = 0.6$. The solid line is the estimate
      due to analytic approches
       and the triangles represent the estimates obtained from
      numerical data.}             
         
   \end{center}
   \end{figure}

The numerical data agrees best with the analytical estimates when
 $\alpha$ is smallest (see figure 6). For this profile we can see 
that the two approaches
agree in the middle of the frequency range, but disagree at low
and high frequencies. The differences at low frequencies are due to
limitations in the procedure outined in equations (38)-(40).
The differences at high frequencies are due to the fact that the
analytic approach is only applicable when the driving frequency is
well below the critical value (equation 257), i.e the linear damping 
approximation.
The difference between numerical and analytic data increases as we
increase the value of $\alpha$ {\change (figures 6, 7, 8 and 9)}. By increasing
this power in the magnetic field profile (equation 25), we are
decreasing the effect of ion-neutral collsional friction
damping, as $\eta_{c}$ is now smaller. This increases the relative importance
of any other effects on the wave amplitude. 
Thus the difference in numerical and analytic data is due to the fact
that the numerical {\change approach} includes the {\change stratification} effects on the
wave amplitudes, whereas the analytic approach does not.

From both analytical and numerical data we can see 
that Alfv\'{e}n waves with frequencies below 0.01 Hz
are unaffected by this
damping mechanism, and propagate through the partially ionised region
with little diffusion. Also waves with frequency above {\change 0.6 Hz} are
completely damped by this mechanism.

%%%%%%%%%%%%%%%%%%%%%%%%%%%%%%%%%%%%%%%%%%%%%%%%%%%%%%%%%%%%%%%%%%%%%%%%%%%%%
%%%%%%%%%%%%%%%%%%%%%%%%%%%%%%%%%%%%%%%%%%%%%%%%%%%%%%%%%%%%%%%%%%%%%%%%%%%%%
\section{Conclusions}

We have estimated the efficiency of the partially ionised
layer of the solar chromosphere in the damping of Alfv\'{e}n waves generated at the
surface. The estimates were based on analytic and numerical
approaches, which agreed in the linear damping approximation.
The damping mechanism was the collisional friction between neutral and
ion species in a model Hydrogen plasma with temperature and density
values taken from the VAL C model of the quiet Sun (\cite{vernazza}).

Alfv\'{e}n waves of frequencies above {\change 0.6 Hz} were completely
damped by the partially ionised layer in the chromosphere, whereas 
waves of frequency
below 0.01 Hz were unaffected by the presence of neutrals and
experienced no damping due to this mechanism.
This lower result agrees with work conducted on MBP's and
power spectra of horizontal motions at different heights in the
atmosphere (\cite{cranmer}). They showed that using the WKB
approximation the power spectra for motions with frequencies below
0.01 Hz are essentially unchanged as one progresses
up the chromosphere to the transition region. This suggests that there
is very little damping of low frequency waves due to any kind of
dissipation mechanism in the solar chromosphere.

Although higher frequency waves are difficult to observe
directly, photospheric
motions can theoretically generate a large spectrum of Alfv\'{e}n waves. The fact that the
partially ionised layer completely damps any waves above {\change 0.6 Hz} for
our magnetic field models
means that high frequency waves in the upper atmosphere must have been
created by other sources than photospheric motions.

{\change This work is based on upward travelling waves generated at the
photospheric level. Downward travelling waves from the corona would be reflected at
the density contrast above the chromospheric region where the Cowling
conductivity is large and it is unclear whether this damping mechanism
would be important. The case of downward propagating waves will be
subject to further investigation.}

The numerical simulations were performed in the non-ideal MHD
approximation with an additional term relating to the Cowling
conductivity in the generalised Ohm's law (equation 16).
Previous work on Alfv\'{e}n wave propagation in the lower solar
atmosphere has used the
WKB assumption, as used by (\cite{depontieu2}). This assumes that the
change in wavelength, as well as variables, is small over a typical
wavelength. They estimated that in this regime the ion-neutral
collisions in the chromosphere damped waves of frequencies above 0.1
Hz.
Our results are in broad agrement with {\change the results of De
  Pontieu et al (2001)} but as is clear
from {\change figures 7,8 and 9} the result is actually sensitive to the spreading
out of the flux tube. Over the range of values of $\alpha$, chosen to
match estimates of $|\mathbf{B}|$ from observations, the damping is
effective over a range 0.1 to {\change 0.6} depending on $\alpha$.

{\change
The model atmosphere was assumed to be magnetised, and thus the
Generalised Ohm's law did not include the Hall term. In the upper
photosphere, the electrons are tightly bound to the magnetic field
whereas the ions are not. In this case there is a seperation electric
field due to the neutrals drag on the non-magnetised ion, which should
be taken into account in the one-fluid MHD equations. However, we
showed that the damping of Alfv\'{e}n waves is efficient at heights of
1000km to 2000km and in this region the plasma can be regarded as magnetised.}
 
The form of the dissipation efficiency calculated from numerical data
differed from that obtained from linear analytic appoximations. 
Small differences could be seen at low frequencies
(due to errors in the estimation of small damping
decrements). Differences also occured at higher frequencies, although 
both analytic and numerical estimates of the efficiency must
converge to 1 at high frequencies. the difference at higher
frequencies is due to  the fact that the linear damping approximation
is not valid (equation 27).

\end{document}